\documentclass[10pt,aps,prl,twocolumn,showpacs,superscriptaddress]{revtex4-1}  

\usepackage[modulo,switch]{lineno}
\modulolinenumbers[1]

\usepackage[utf8]{inputenc}
\usepackage{graphicx}
\usepackage{amsmath}
\usepackage{amssymb}
\usepackage{bm}        
\usepackage{txfonts}
\usepackage{multirow}
\usepackage{mathptmx}  
\usepackage{siunitx}

\usepackage{xargs}                      

\usepackage[%
  colorlinks=true,
  urlcolor=blue,
  linkcolor=blue,
  citecolor=blue
]{hyperref}

\begin{document}

\preprint{AIP/123-QED}

\title[Motion analysis of a trapped ion chain \ldots]{Motion analysis of a trapped ion chain by single photon self-interference}

\author{G.~Cerchiari}
\email{giovanni.cerchiari@uibk.ac.at}
\affiliation{Institut f\"ur Experimentalphysik, Universit\"at Innsbruck, Technikerstrasse~25, 6020~Innsbruck, Austria}
\author{G.~Araneda}
\affiliation{Institut f\"ur Experimentalphysik, Universit\"at Innsbruck, Technikerstrasse~25, 6020~Innsbruck, Austria}
\affiliation{Clarendon Laboratory, Department of Physics, University of Oxford, Parks Road, Oxford OX1 3PU, United Kingdom}
\author{L.~Podhora}
\affiliation{Department of Optics, Palack\'y University, 17. Listopadu 12, 77146 Olomouc, Czech Republic}
\author{L.~Slodi\v{c}ka}
\affiliation{Department of Optics, Palack\'y University, 17. Listopadu 12, 77146 Olomouc, Czech Republic}
\author{Y.~Colombe}
\affiliation{Institut f\"ur Experimentalphysik, Universit\"at Innsbruck, Technikerstrasse~25, 6020~Innsbruck, Austria}
\author{R.~Blatt}
\affiliation{Institut f\"ur Experimentalphysik, Universit\"at Innsbruck, Technikerstrasse~25, 6020~Innsbruck, Austria}
\affiliation{Institut f\"ur Quantenoptik und Quanteninformation, \"Osterreichische Akademie der Wissenschaften, Technikerstrasse 21a, 6020 Innsbruck, Austria}

\date{\today}

\begin{abstract}
We present an optical scheme to detect the oscillations of a two-ion string confined in a linear Paul trap. The motion is detected by analyzing the intensity correlations in the fluorescence light emitted by one or two ions in the string. We present measurements performed under continuous Doppler cooling and under pulsed illumination. We foresee several direct applications of this detection method, including motional analysis of multi-ion species or coupled mechanical oscillators, and sensing of mechanical correlations.
\end{abstract}

\maketitle

%

Observation of the oscillations of trapped ions is an essential technique used in cutting-edge quantum~\cite{leibfried2003quantum} and fundamental~\cite{FundamentalPhysicsinParticleTraps} physics experiments. The estimations of bare oscillation frequencies are used to provide precise values of the rest energies~\cite{Schuessler2020} of atoms playing a key role for the estimation of the neutrino mass~\cite{PhysRevLett.115.062501}. Differences in oscillation frequencies are also investigated in precision spectroscopy experiments~\cite{Egl2019} to measure the gyromagnetic factors of fundamental particles, which is relevant for tests of QED~\cite{PhysRevLett.107.023002} and to search for asymmetry between matter and antimatter~\cite{Nature24048}. The conventional approach in Penning traps is to detect the current induced by the ion's image charge on the trapping electrodes~\cite{FundamentalPhysicsinParticleTraps}. Novel methods are being explored to perform precision measurements with higher sensitivity to motion by using a second ion~\cite{Repp2012}. The additional ion should have a favorable electronic structure to prepare and read-out the ion's property of interest via quantum logic spectroscopy~\cite{PhysRevA.85.043412}. The quantum logic scheme requires several well-controlled laser pulses to manipulate the auxiliary ion. This ion is first prepared by laser cooling and then interrogated by addressing the motional sidebands of the chain using a narrow transition. 

Techniques alternative to sideband spectroscopy for motion detection have been explored in past which rely on the analysis of the scattered light~\cite{dholakia1993photon, Raab2000, Bushev2006, rotter2008monitoring, Bushev2013}. These techniques are based on the self-interference of the single photons emitted by an atom on a dipole transition. By improving on this technique, we recently demonstrated that the self-interference of single photons can be used to measure the oscillations of a trapped ion down to single quanta of oscillation~\cite{Cerchiari2020}. This technique can present advantages over quantum logic spectroscopy. First, it allows measuring all the oscillation modes simultaneously without prior knowledge of their frequencies. Second, it requires only the laser radiation suitable for Doppler cooling of the ion and, in general, it is suited for detecting the motion of any dipolar scatterer\cite{Cerchiari2021position}.

In this article, we describe how this method can be extended to the study of the oscillations of a chain of two atomic ions. The detection can be performed by observing the fluorescence of a single ion and thus offers interesting prospects for the study of chains containing multiple ion species. First, we summarize the experimental setup for motion detection. Then, we describe the measurements of the chain oscillations under continuous Doppler cooling conditions or performed with pulsed excitation. We present the two regimes because they can be suited or combined in a rich variety of applications for motional analysis. For example, they can be used to estimate the mechanical frequency oscillation of massive silicon nanoparticles or for the coherent analysis of the motion of trapped atoms or molecules. The continuous measurement can be used for detecting all the mechanical modes simultaneously in short time, and the pulsed excitation scheme is preferred for applications sensitive to photon recoil, including the phase measurement of coherent oscillation or the population analysis of quantum states near the motional ground state. \\

\begin{figure*}
\centerline{\includegraphics[width=0.9\textwidth]{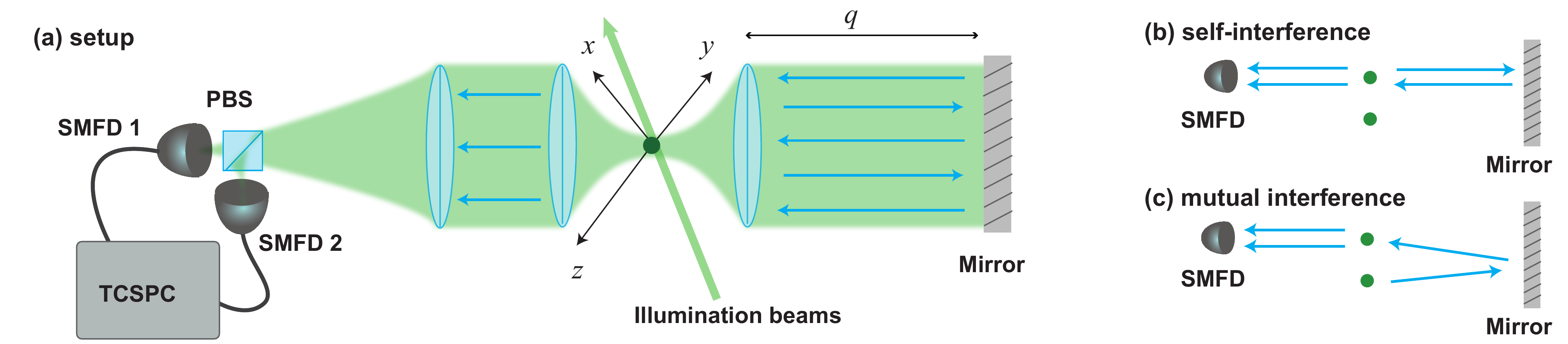}}
\caption{(a) A single ion or a linear chain  of two atomic ions aligned along the z-axis is confined in a Paul trap. The system exhibits three normal oscillation directions, $x$, $y$, and $z$. The chain is illuminated with laser beams and part of the scattered light is collected by two confocal lenses. A mirror is used to overlap the light emitted to create a standing wave. A polarizing beam splitter is used to separate different linear polarizations and the light is then coupled to single-mode fiber detectors (SMFD 1 and SMFD 2). Each detected photon is time-tagged using a Time-Correlated Single Photon Counting system (TCSPC). The detection configurations for multiple trapped ions are presented in subfigures (b) and (c). (b) ``Self'' configuration: Part of the light emitted by a single ion is reflected back on the same ion and subsequently detected. (c) ``Mutual'' configuration: The light emitted by one ion is reflected and superimposed with the light emitted by the second ion. For simplicity, only one single-mode-fiber photon detector (SMFD) is depicted in each configuration.}
\label{fig:setup}
\end{figure*}

In the experiments, we cool and trap $^{138}$Ba$^+$ ions in a linear Paul trap driven at the frequency $\Omega_{\textrm{RF}}=2\pi\times15$~MHz. The motion of a single ion can be approximated as harmonic with three normal modes in three orthogonal directions. One mode is aligned with the trap axis ($z$-direction) and has a frequency lower than the other two: $\omega_{x1} \simeq \omega_{y1} > \omega_{z1}$. The frequencies span the interval between $2\pi\cdot 0.5$~MHz and $2\pi\cdot 1.8$~MHz. For co-trapped ions, the motion is coupled by the Coulomb interaction. For each normal direction ($x$, $y$, $z$) the chain has two modes of oscillation~\cite{Home_2011}: one where the ions move in phase, the common mode, and one where the ions move out of phase, the rocking or breathing mode. For the used trapping frequencies and at the Doppler limit, the peak-to-peak amplitude of the oscillation of the atoms is $\sim80$~nm in the axial center of mass (COM) mode and $\sim40$~nm in the other modes, calculated from the theoretical Doppler limit temperature at our laser parameters.

The setup and the laser parameters are described thoroughly in Ref.~\onlinecite{Cerchiari2020} in the context of measurements with a single ion. Here we summarize the characteristics relevant for the presented measurements. The laser beam used for Doppler cooling propagates across the trap at an angle of 45$^\circ$ with respect to the trap axis ($z$-axis) to ensure cooling of all the oscillation modes. Two lasers drive the lambda transitions $6\textrm{S}_{1/2} \leftrightarrow 6\textrm{P}_{1/2}\leftrightarrow 5\textrm{D}_{3/2}$. The fluorescence light of the $6\textrm{P}_{1/2} \rightarrow 6\textrm{S}_{1/2}$ transition is collimated by a pair of confocal objectives ($\textrm{NA}\sim 0.40$), see Fig.~\ref{fig:setup}(a). On one side, at a distance $q\sim30$~cm from the ions, the light is back-reflected by a mirror. On the opposite side, the back-reflected fluorescence and direct fluorescence are collimated by the second objective towards the detectors. The detectors are two single-mode fiber-coupled single-photon avalanche photodetectors (APDs) connected to a Time-Correlated Single Photon Counting system (TCSPC, PicoHarp 300) to time-tag the arrival time of each photon. We combine the information of the two detectors as if it would be only one. The two detectors in Hanbury Brown and Twiss configuration are not necessary for these experiments, because we are not interested in correlating photons with time separation below the dead time of the APDs ($\sim100$~ns). We adopted this configuration because it was already prepared in the laboratory during previous measurements~\cite{Araneda_2018}.

The detection rate $R$ on the detectors depends on the interference of single fluorescence photons. The rate $R$ can be modeled as the interference between the direct fluorescence and reflected fluorescence at the detector. The rate follows a sinusoidal dependence as a function of the distance $q$ between the detected ion and the mirror:
\begin{equation} \label{eq_fringe}
R= R_0 \left(1 + \nu \sin\left(\frac{4\pi}{\lambda}q\right)\right)\, ,
\end{equation}
where the rate $R_0 \sim 5\cdot10^4$ photons/s is stable on the time scale of the measurements ($\sim1000$~s). The contrast of the interference fringes ranging from $\nu \sim 30\%$ to $40\%$ has been measured by scanning the relative distance $q$ between the ion and the mirror. The fluorescence rate is modulated around $1$~MHz by the ion oscillations. Slow variations ($\sim1$~s) are caused by drifts in the mirror position. The average fluorescence rate over 100~ms is monitored to correct the mirror position in a feedback loop to keep the interference condition on the slope of the fringe. The ions are illuminated continuously with the Doppler cooling beam while the photon events are time-tagged by the detection system. The total acquisition time corresponding to the presented measurements was 900~s. After this time, the acquisition is saved and post-processed to analyze the oscillations. We use a feedback circuit to stabilize the RF drive amplitude and increase the apparent coherence time of the oscillations because fluctuations in the driving field power can induce variations in the strength of the harmonic confinement. More details on the RF stabilization can be found in Ref.~\onlinecite{Cerchiari2020}. \\

\begin{figure*}
\centerline{\includegraphics[width=1\textwidth]{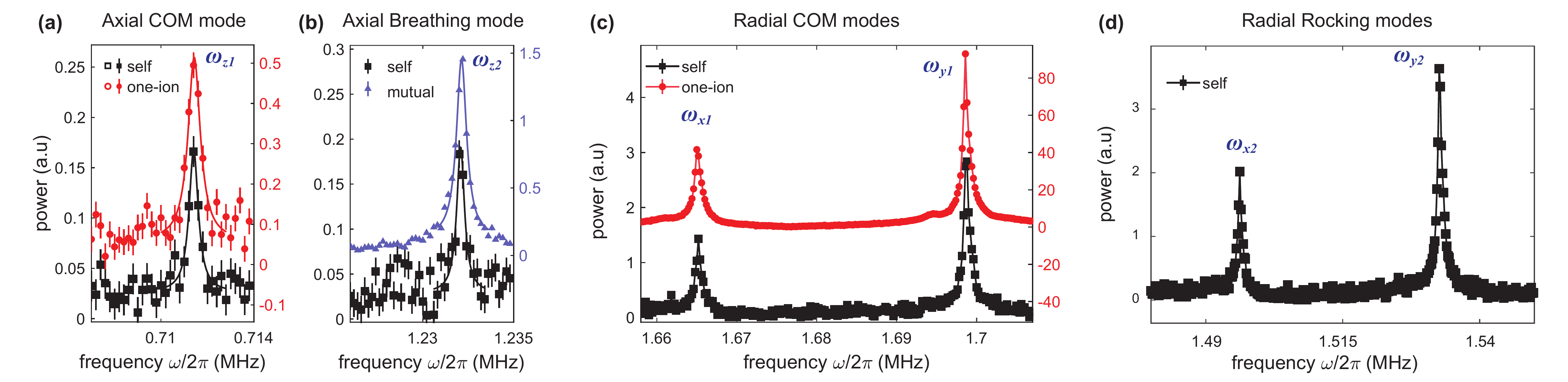}}
\caption{Examples of motional power spectra of one- or two-ion chains. (a) Axial center of mass mode. (b) Axial breathing mode. (c) Radial center of mass (COM) modes. (d) Radial rocking modes. Note the different vertical scales (color-coded). The label ``self'' and ``mutual'' refers to the ``self-interference'' and ``mutual interference'' configurations described in Fig.~\ref{fig:setup} (b) and (c). The lines in (a) and (b) are Lorentian fit to the data obtained as described in the main text. The lines in (c) and (d) are meant to guide the eye.}
\label{fig:continous}
\end{figure*}

The continuous measurements with two trapped ions correspond to the conventional excitation regime with a close-to Doppler cooling limit of motional energies. The advantages of this regime include the simplicity of the scheme and the amount of the signal acquired over the finite measurement time. However, at the same time, the analyzed particle motion is forced to follow the laser cooling dynamics corresponding to random photon recoils which result in the thermal steady state. For experiments with two trapped ions presented here, we employ two configurations of alignment of the optical interference setup. The ``self-interference'' configuration is depicted in Fig.~\ref{fig:setup}(b) and results in the interference of one ion with its mirror image. Alternatively, the ``mutual interference'' configuration depicted in Fig.~\ref{fig:setup}(b) combines the light from the two ions on the detector.

For experiments in the ``self-interference'' configuration the fluorescence of the second ion is not detected. The cross-talk from the second ion on the detector is negligible thanks to the large numerical aperture $\textrm{NA}\sim0.4$ of the confocal objectives and thanks to the single-mode fibers guiding the light to the APDs. The motion is analyzed by calculating the Fourier transform of the two-photon correlation function $g^2(\tau)$. The correlation is performed with a histogram of (8 to 32)~ns bin size on an interval of 1 to 10~ms. Typical spectra calculated with a maximum correlation time of 5~ms are presented in Fig.~\ref{fig:continous} for a single ion or a string of two ions. The ``one-ion'' spectrum, with measured data shown as red circles, corresponds to the measurement of a single oscillating ion in the trap. The data points shown as black squares are the spectrum obtained in the ``self-interference'' configuration with two ions. The detection scheme does not require prior knowledge of the system characteristic frequencies and reveals all the oscillations modes simultaneously.

\begin{figure}
\centerline{\includegraphics[scale=0.55]{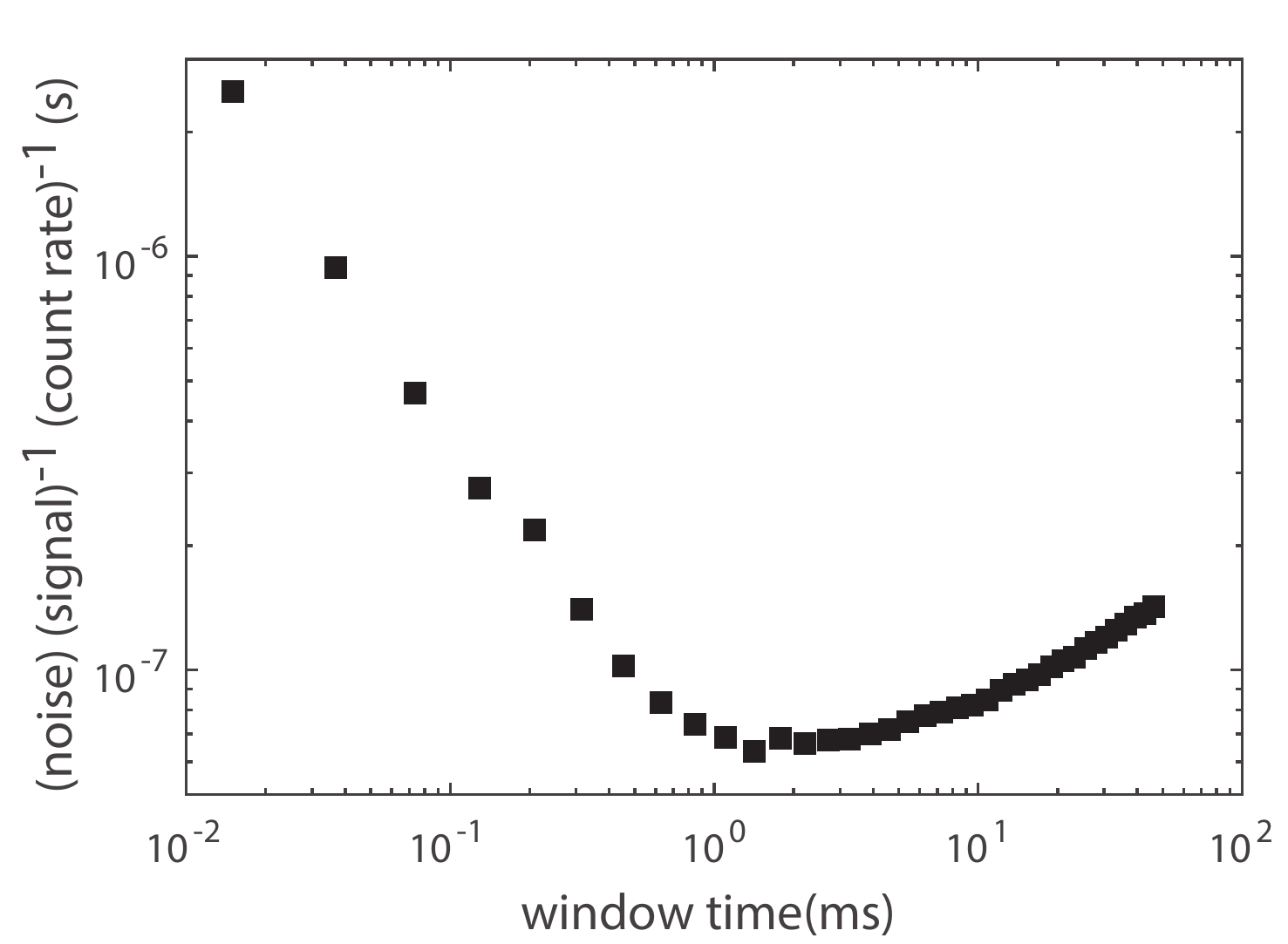}}
\caption{Noise-to-signal ratio as a function of the correlation time window. See details in main text.}
\label{fig:noise_to_signal}
\end{figure}

The optimal correlation length is calculated from the observed noise-to-signal ratio for different time windows. In this analysis, the signal is the amplitude of the highest peak in the spectrum and the noise is the standard deviation of the amplitudes in the frequency range 2 to 3~MHz, where no peak is present. Fig.~\ref{fig:noise_to_signal} shows the noise-to-signal ratio for the peak $\omega_{x1}$ for different maximum correlation time, in the case of a single trapped ion. The highest sensitivity is obtained for a histogram with maximum correlation time of 1 to 5~ms. This time is mostly determined by the observed coherence of the oscillation, which is limited by the stability of the RF electric field's power driving the Paul trap~\cite{Cerchiari2020}. This decoherence is visible in the correlation function as an exponential damping of the oscillations which emerges from averaging the correlation over all the photon pairs. For our acquisition time of $\sim900$~s, the damped oscillations become comparable to the noise fluctuations at correlation times of a few ms, thus determining the optimal maximum correlation time.

\begin{table}
 \vspace{-\abovecaptionskip}
 \caption{Measured oscillation frequencies for the data reported in Fig. \ref{fig:setup}.}
 \label{tab:frequency}
\vspace{\abovecaptionskip}
\centering
\begin{tabular}{ l  c  c }
 \hline\hline
mode &one ion (MHz) & two ions (MHz)  \\
 \hline  \hline
 $\omega_{z1}/2\pi$ & $0.711429(22)$ & $0.711407(23)$  \\
 \hline
 $\omega_{z2}/2\pi$ & n.a. & $1.232080(27)$  \\
 \hline
 $\omega_{x2}/2\pi$ & n.a. & $1.496213(20)$  \\
 \hline
 $\omega_{y2}/2\pi$ & n.a. & $1.532719(21)$  \\
 \hline
 $\omega_{x1}/2\pi$ & $1.665117(15)$ & $1.665212(19)$  \\
 \hline
 $\omega_{y1}/2\pi$ & $1.698637(21)$ & $1.698744(21)$  \\
 \hline\hline
\end{tabular}
\end{table}

Taking advantage of the fact that we are working with identical ions, we can also superimpose the fluorescence of both ions onto the detectors. To do so, we rotate the mirror to direct the reflected fluorescence of one ion onto the second ion, as depicted in the ``mutual interference'' configuration in Fig.~\ref{fig:setup}(c). As for the previous measurement performed in the ``self-interference'' configuration (Fig.~\ref{fig:setup}(b)), the characteristics of the optical setup exclude any cross-talk also in the ``mutual interference'' configuration. Compared to the acquisition in the ``self-interference'' arrangement, the ``mutual interference'' configuration suffers from a reduced interference contrast due to the sensitivity to the ratio of the elastically to inelastically scattered part of the fluorescence. Thus, the ``mutual interference'' configuration is less suitable for the sensing of mechanical oscillations with the exception of the axial breathing mode, whose data are presented in Fig.~\ref{fig:continous}(b). The higher sensitivity on this mode can be understood by observing that the relative distance of the two ions in the ``mutual interference'' configuration changes the relative path length in the ion-mirror interferometer as reported in Ref.~\onlinecite{Araneda_2018}. Further data that compare the sensitivity of the two configurations are presented in the supplementary material.

The characteristic oscillation frequencies of the single-atom and the two-atom string can be identified. To estimate the mode frequencies, each peak was isolated in a frequency window of 5~kHz and a Lorentzian resonance curve was fitted to the data. Typical linewidths of the peaks extracted from the Lorentzian fit are in the range 700 to 900~Hz. The oscillation frequencies of the shown data are reported in Tab.~\ref{tab:frequency}. Our detection scheme is mostly sensitive to motion with projection along the optical axis ($x$ and $y$ radial modes), but can also identify oscillations in the mode orthogonal to the optical axis ($z$-mode). The latter is measurable because of the non-perfect orthogonal alignment of the detection axis with respect to the trap axis and because the ion moves across the optical mode collected by the objectives. The frequencies can be identified down to a 20 to 30~Hz uncertainty from the Lorentzian fits of the peaks in the spectrum.

The circuit that stabilizes the RF power increases the quality factor but is known to introduce non-linearities~\cite{Cerchiari2020}. The non-linear effect can be identified in the ``one-ion'' data as a kink around $1.695$~MHz and is below the noise floor in the ``self-interference'' data.\\

While the motional detection in the continuous excitation regime allows for the precise and simultaneous detection of motional frequencies of atoms, it also presents some fundamental limitations when considering light atomic particles with internal electronic level structure. First, the scattered light induces a back-action which may reduce the coherence of the oscillator. In our experiment, this effect could not be measured because the decoherence caused by back-action is expected to appear at the time scale of $\sim2$~ms\cite{Cerchiari2020}, while the coherence time of the oscillations is $\sim800~\mu$s. The observed coherence time is limited by the stability of the trapping field's power and results from averaging over multiple runs of the experiment. Second, the presence of the mirror modifies the energy level of the excited state of the atom. The energy variation follows the sinusoidal pattern of Eq.~\ref{eq_fringe}, but it is phase-shifted by $\pi/2$~\cite{Dorner2002, Hetet2010}. The energy shift is relevant only for atomic and molecular systems and may cause stronger or weaker confinement of the ion in the trap, which manifests itself as a deviation in the measured trapping frequency. For example, in Ref.~\onlinecite{Bushev_2004} a peak-to-peak deviation of 310~Hz for a trapped Ba$^+$ oscillating at 1~MHz with an optical setup having $\textrm{NA}\sim0.4$ was reported. Furthermore, the sign of the shift depends on the derivative of the slope of the interference fringe. A possible strategy for suppression of this systematic frequency deviation would employ averaging the frequency measurements over the two mirror positions at opposite slopes of the interference fringe. These interference points correspond to substitute the factor $\nu$ with $+\nu$ and $-\nu$ in Eq.~\ref{eq_fringe}.

\begin{figure}
\centering
\includegraphics[width=1\columnwidth]{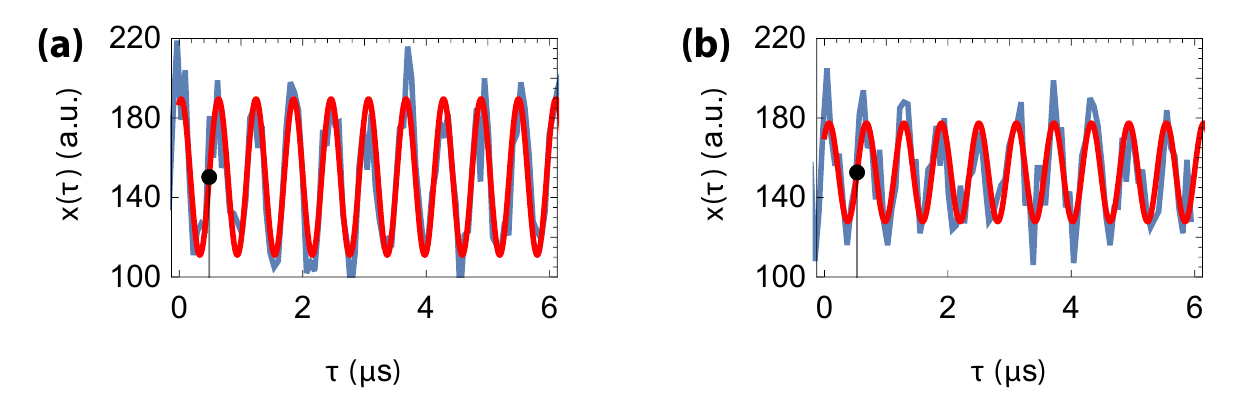}
\caption{Phase-sensitive measurement. The ion oscillations are probed by pulsed illumination during a short $\sim6\,\mu$s interval. The ion is not illuminated at other times. (a) Histogram of the photons arriving after $t=62\,\mu$s ($\tau=0$ in the plot) relative to the start of the amplification drive. (b) Same plot for $t=413\,\mu$s. The points $t_1$ and $t_2$, marked with black vertical lines, have the same phase and are discussed in the main text.}
\label{fig:2_oscillation_phase_points}
\end{figure}

The back-action and the systematic shift are induced by the light scattering process. For this reason, we explore an alternative excitation regime in which the ion can be kept unilluminated for most of the measurement cycle. This regime consists of the initial period in which the ion is excited by an external RF electric field resonant with a particular motional mode of interest. In our measurement, at time $t=0$ an RF pulse of 50~$\mu$s is applied with a known phase to excite the $x$-mode. The drive increases the peak-to-peak amplitude of the ion chain motion to $\sim100$~nm, with a well-defined phase. Thereafter, the apparent amplitude decreases with a time constant of $0.85$~ms. We record the scattered photons during different specific time windows, switching on the cooling beam for a limited amount of time. We record approximately 20 oscillations starting alternatively around $t\sim62\,\mu$s or $t\sim413\,\mu$s after the trigger of the driving pulse. The experiment is repeated for 900~s, and the detected signal is post-processed.

The start of the drive is used as a time reference for the histogram of the photon events (see Fig. \ref{fig:2_oscillation_phase_points}). Each signal is fitted with a sinusoidal oscillation to extract the phase at a specific elapsed time. From the fit we find that the same phase repeats at times $t_{1} = 62.484(5) \,\mu $s and $t_{2} = 413.528(8) \, \mu$s. Considering the number of oscillations between the two acquisitions ($n = 578$) we estimate a frequency of oscillation $\omega_{x1}/2\pi = 1.64652(4) $~MHz. The frequency uncertainty has the same order of magnitude as achieved with the method presented earlier in the article, showing that this uncertainty is dominated by the instabilities of the trapping field rather than by the decoherence induced by the back-action of the cooling light. The values measured in the two sections differ by a few kHz, which can be mostly attributed to the drift of the trapping frequency on the long time scales between the experimental runs.\\

In this article, we measured the mechanical oscillations of a two-ion string from the interference of single photons emitted by the ions. The method is based on the interference of the light scattered by the trapped particle and can be applied to a large range of oscillators, including levitated nanoparticles or long ion chains~\cite{obvsil2019multipath}. 

We presented the measurement results obtained with a two-ion chain oscillating close to the Doppler cooling limit. The entire set of oscillation modes can be monitored simultaneously by detecting the fluorescence field emitted by one or both ions. In addition, we implemented a pulsed measurement aimed to overcome the limitations induced by continuous excitation of atoms, including the back-action and energy level shifts. The oscillations can be probed by imprinting a phase with an external drive and using delayed pulsed illumination to observe the oscillation at two distant time points. 

The method can be applied by utilizing the ''self-interference'' of single photons emitted by an ion or the ''mutual interference'' between the fluorescence of two ions. The ''self-interference'', in which one ion only is monitored, has applications in the study of the oscillation of hybrid chains of trapped particles that include, for example, ions without an accessible Doppler cooling transition. The sensitivity of the scheme could be widely improved by using high numerical aperture systems and high self-interference stability, as the one presented in Ref.~\onlinecite{araneda2020panopticon}. The phase-sensitive technique in combination with EIT cooling\cite{lechner2016electromagnetically} could be used to prepare and measure the oscillations of the ion chain close to the motional ground state without the need of additional narrow transitions typically used for these purposes.

\section{acknowledgments}

We thank Pavel Bushev, Alexander Rischka and Dougal Main for insightful discussions.
L.P. and L.S. acknowledge the support of the Czech Science Foundation under Grant No. GA19-14988S. This work was supported by the European Union through Horizon 2020 grant agreement No. 801285 (PIEDMONS project) and by the Institut für Quanteninformation GmbH. This work has also received funding from the European Union’s Horizon 2020 research and innovation program under the Marie Skłodowska-Curie grant agreement No. 801110 and the Austrian Federal Ministry of Education, Science and Research (BMBWF). This article reflects only the authors’ view and the European Union cannot be held responsible for any use that may be made of the information it contains. This work has received support from the Institut für Quanteninformation GmbH.

\bibliography{bibliography}

\begin{thebibliography}{24}
\expandafter\ifx\csname natexlab\endcsname\relax\def\natexlab#1{#1}\fi
\expandafter\ifx\csname bibnamefont\endcsname\relax
  \def\bibnamefont#1{#1}\fi
\expandafter\ifx\csname bibfnamefont\endcsname\relax
  \def\bibfnamefont#1{#1}\fi
\expandafter\ifx\csname citenamefont\endcsname\relax
  \def\citenamefont#1{#1}\fi
\expandafter\ifx\csname url\endcsname\relax
  \def\url#1{\texttt{#1}}\fi
\expandafter\ifx\csname urlprefix\endcsname\relax\def\urlprefix{URL }\fi
\providecommand{\bibinfo}[2]{#2}
\providecommand{\eprint}[2][]{\url{#2}}

\bibitem[{\citenamefont{Leibfried et~al.}(2003)\citenamefont{Leibfried, Blatt,
  Monroe, and Wineland}}]{leibfried2003quantum}
\bibinfo{author}{\bibfnamefont{D.}~\bibnamefont{Leibfried}},
  \bibinfo{author}{\bibfnamefont{R.}~\bibnamefont{Blatt}},
  \bibinfo{author}{\bibfnamefont{C.}~\bibnamefont{Monroe}}, \bibnamefont{and}
  \bibinfo{author}{\bibfnamefont{D.}~\bibnamefont{Wineland}},
  \bibinfo{journal}{Rev. Mod. Phys.} \textbf{\bibinfo{volume}{75}},
  \bibinfo{pages}{281} (\bibinfo{year}{2003}),
  \urlprefix\url{https://link.aps.org/doi/10.1103/RevModPhys.75.281}.

\bibitem[{\citenamefont{Quint~W.}(2014)}]{FundamentalPhysicsinParticleTraps}
\bibinfo{editor}{\bibfnamefont{V.~M.} \bibnamefont{Quint~W.}}, ed.,
  \emph{\bibinfo{title}{Fundamental Physics in Particle Traps}}
  (\bibinfo{publisher}{Springer}, \bibinfo{year}{2014}), ISBN
  \bibinfo{isbn}{978-3-662-51173-2},
  \urlprefix\url{https://doi.org/10.1007/978-3-642-45201-7}.

\bibitem[{\citenamefont{Schüssler et~al.}(2020)\citenamefont{Schüssler,
  Bekker, Braß, Cakir, Crespo López-Urrutia, Door, Filianin, Harman,
  Haverkort, Huang et~al.}}]{Schuessler2020}
\bibinfo{author}{\bibfnamefont{R.~X.} \bibnamefont{Schüssler}},
  \bibinfo{author}{\bibfnamefont{H.}~\bibnamefont{Bekker}},
  \bibinfo{author}{\bibfnamefont{M.}~\bibnamefont{Braß}},
  \bibinfo{author}{\bibfnamefont{H.}~\bibnamefont{Cakir}},
  \bibinfo{author}{\bibfnamefont{J.~R.} \bibnamefont{Crespo López-Urrutia}},
  \bibinfo{author}{\bibfnamefont{M.}~\bibnamefont{Door}},
  \bibinfo{author}{\bibfnamefont{P.}~\bibnamefont{Filianin}},
  \bibinfo{author}{\bibfnamefont{Z.}~\bibnamefont{Harman}},
  \bibinfo{author}{\bibfnamefont{M.~W.} \bibnamefont{Haverkort}},
  \bibinfo{author}{\bibfnamefont{W.~J.} \bibnamefont{Huang}},
  \bibnamefont{et~al.}, \bibinfo{journal}{Nature}
  \textbf{\bibinfo{volume}{581}}, \bibinfo{pages}{42} (\bibinfo{year}{2020}),
  ISSN \bibinfo{issn}{1476-4687},
  \urlprefix\url{https://doi.org/10.1038/s41586-020-2221-0}.

\bibitem[{\citenamefont{Eliseev et~al.}(2015)\citenamefont{Eliseev, Blaum,
  Block, Chenmarev, Dorrer, D\"ullmann, Enss, Filianin, Gastaldo, Goncharov
  et~al.}}]{PhysRevLett.115.062501}
\bibinfo{author}{\bibfnamefont{S.}~\bibnamefont{Eliseev}},
  \bibinfo{author}{\bibfnamefont{K.}~\bibnamefont{Blaum}},
  \bibinfo{author}{\bibfnamefont{M.}~\bibnamefont{Block}},
  \bibinfo{author}{\bibfnamefont{S.}~\bibnamefont{Chenmarev}},
  \bibinfo{author}{\bibfnamefont{H.}~\bibnamefont{Dorrer}},
  \bibinfo{author}{\bibfnamefont{C.~E.} \bibnamefont{D\"ullmann}},
  \bibinfo{author}{\bibfnamefont{C.}~\bibnamefont{Enss}},
  \bibinfo{author}{\bibfnamefont{P.~E.} \bibnamefont{Filianin}},
  \bibinfo{author}{\bibfnamefont{L.}~\bibnamefont{Gastaldo}},
  \bibinfo{author}{\bibfnamefont{M.}~\bibnamefont{Goncharov}},
  \bibnamefont{et~al.}, \bibinfo{journal}{Phys. Rev. Lett.}
  \textbf{\bibinfo{volume}{115}}, \bibinfo{pages}{062501}
  (\bibinfo{year}{2015}),
  \urlprefix\url{https://link.aps.org/doi/10.1103/PhysRevLett.115.062501}.

\bibitem[{\citenamefont{Egl et~al.}(2019)\citenamefont{Egl, Arapoglou,
  H\"ocker, K\"onig, Ratajczyk, Sailer, Tu, Weigel, Blaum, N\"ortersh\"auser
  et~al.}}]{Egl2019}
\bibinfo{author}{\bibfnamefont{A.}~\bibnamefont{Egl}},
  \bibinfo{author}{\bibfnamefont{I.}~\bibnamefont{Arapoglou}},
  \bibinfo{author}{\bibfnamefont{M.}~\bibnamefont{H\"ocker}},
  \bibinfo{author}{\bibfnamefont{K.}~\bibnamefont{K\"onig}},
  \bibinfo{author}{\bibfnamefont{T.}~\bibnamefont{Ratajczyk}},
  \bibinfo{author}{\bibfnamefont{T.}~\bibnamefont{Sailer}},
  \bibinfo{author}{\bibfnamefont{B.}~\bibnamefont{Tu}},
  \bibinfo{author}{\bibfnamefont{A.}~\bibnamefont{Weigel}},
  \bibinfo{author}{\bibfnamefont{K.}~\bibnamefont{Blaum}},
  \bibinfo{author}{\bibfnamefont{W.}~\bibnamefont{N\"ortersh\"auser}},
  \bibnamefont{et~al.}, \bibinfo{journal}{Phys. Rev. Lett.}
  \textbf{\bibinfo{volume}{123}}, \bibinfo{pages}{123001}
  (\bibinfo{year}{2019}),
  \urlprefix\url{https://link.aps.org/doi/10.1103/PhysRevLett.123.123001}.

\bibitem[{\citenamefont{Sturm et~al.}(2011)\citenamefont{Sturm, Wagner,
  Schabinger, Zatorski, Harman, Quint, Werth, Keitel, and
  Blaum}}]{PhysRevLett.107.023002}
\bibinfo{author}{\bibfnamefont{S.}~\bibnamefont{Sturm}},
  \bibinfo{author}{\bibfnamefont{A.}~\bibnamefont{Wagner}},
  \bibinfo{author}{\bibfnamefont{B.}~\bibnamefont{Schabinger}},
  \bibinfo{author}{\bibfnamefont{J.}~\bibnamefont{Zatorski}},
  \bibinfo{author}{\bibfnamefont{Z.}~\bibnamefont{Harman}},
  \bibinfo{author}{\bibfnamefont{W.}~\bibnamefont{Quint}},
  \bibinfo{author}{\bibfnamefont{G.}~\bibnamefont{Werth}},
  \bibinfo{author}{\bibfnamefont{C.~H.} \bibnamefont{Keitel}},
  \bibnamefont{and} \bibinfo{author}{\bibfnamefont{K.}~\bibnamefont{Blaum}},
  \bibinfo{journal}{Phys. Rev. Lett.} \textbf{\bibinfo{volume}{107}},
  \bibinfo{pages}{023002} (\bibinfo{year}{2011}),
  \urlprefix\url{https://link.aps.org/doi/10.1103/PhysRevLett.107.023002}.

\bibitem[{\citenamefont{Smorra et~al.}(2017)\citenamefont{Smorra, Sellner,
  Borchert, Harrington, Higuchi, Nagahama, Tanaka, Mooser, Schneider, Bohman
  et~al.}}]{Nature24048}
\bibinfo{author}{\bibfnamefont{C.}~\bibnamefont{Smorra}},
  \bibinfo{author}{\bibfnamefont{S.}~\bibnamefont{Sellner}},
  \bibinfo{author}{\bibfnamefont{M.~J.} \bibnamefont{Borchert}},
  \bibinfo{author}{\bibfnamefont{J.~A.} \bibnamefont{Harrington}},
  \bibinfo{author}{\bibfnamefont{T.}~\bibnamefont{Higuchi}},
  \bibinfo{author}{\bibfnamefont{H.}~\bibnamefont{Nagahama}},
  \bibinfo{author}{\bibfnamefont{T.}~\bibnamefont{Tanaka}},
  \bibinfo{author}{\bibfnamefont{A.}~\bibnamefont{Mooser}},
  \bibinfo{author}{\bibfnamefont{G.}~\bibnamefont{Schneider}},
  \bibinfo{author}{\bibfnamefont{M.}~\bibnamefont{Bohman}},
  \bibnamefont{et~al.}, \bibinfo{journal}{Nature}
  \textbf{\bibinfo{volume}{550}}, \bibinfo{pages}{371} (\bibinfo{year}{2017}),
  \urlprefix\url{https://doi.org/10.1038/nature24048}.

\bibitem[{\citenamefont{Repp et~al.}(2012)\citenamefont{Repp, B{\"o}hm, Crespo
  L{\'o}pez-Urrutia, D{\"o}rr, Eliseev, George, Goncharov, Novikov, Roux, Sturm
  et~al.}}]{Repp2012}
\bibinfo{author}{\bibfnamefont{J.}~\bibnamefont{Repp}},
  \bibinfo{author}{\bibfnamefont{C.}~\bibnamefont{B{\"o}hm}},
  \bibinfo{author}{\bibfnamefont{J.~R.} \bibnamefont{Crespo
  L{\'o}pez-Urrutia}},
  \bibinfo{author}{\bibfnamefont{A.}~\bibnamefont{D{\"o}rr}},
  \bibinfo{author}{\bibfnamefont{S.}~\bibnamefont{Eliseev}},
  \bibinfo{author}{\bibfnamefont{S.}~\bibnamefont{George}},
  \bibinfo{author}{\bibfnamefont{M.}~\bibnamefont{Goncharov}},
  \bibinfo{author}{\bibfnamefont{Y.~N.} \bibnamefont{Novikov}},
  \bibinfo{author}{\bibfnamefont{C.}~\bibnamefont{Roux}},
  \bibinfo{author}{\bibfnamefont{S.}~\bibnamefont{Sturm}},
  \bibnamefont{et~al.}, \bibinfo{journal}{Applied Physics B}
  \textbf{\bibinfo{volume}{107}}, \bibinfo{pages}{983} (\bibinfo{year}{2012}),
  ISSN \bibinfo{issn}{1432-0649},
  \urlprefix\url{https://doi.org/10.1007/s00340-011-4823-6}.

\bibitem[{\citenamefont{W\"ubbena et~al.}(2012)\citenamefont{W\"ubbena, Amairi,
  Mandel, and Schmidt}}]{PhysRevA.85.043412}
\bibinfo{author}{\bibfnamefont{J.~B.} \bibnamefont{W\"ubbena}},
  \bibinfo{author}{\bibfnamefont{S.}~\bibnamefont{Amairi}},
  \bibinfo{author}{\bibfnamefont{O.}~\bibnamefont{Mandel}}, \bibnamefont{and}
  \bibinfo{author}{\bibfnamefont{P.~O.} \bibnamefont{Schmidt}},
  \bibinfo{journal}{Phys. Rev. A} \textbf{\bibinfo{volume}{85}},
  \bibinfo{pages}{043412} (\bibinfo{year}{2012}),
  \urlprefix\url{https://link.aps.org/doi/10.1103/PhysRevA.85.043412}.

\bibitem[{\citenamefont{Dholakia et~al.}(1993)\citenamefont{Dholakia, Horvath,
  Segal, Thompson, Warrington, and Wilson}}]{dholakia1993photon}
\bibinfo{author}{\bibfnamefont{K.}~\bibnamefont{Dholakia}},
  \bibinfo{author}{\bibfnamefont{G.~Z.~K.} \bibnamefont{Horvath}},
  \bibinfo{author}{\bibfnamefont{D.~M.} \bibnamefont{Segal}},
  \bibinfo{author}{\bibfnamefont{R.~C.} \bibnamefont{Thompson}},
  \bibinfo{author}{\bibfnamefont{D.~M.} \bibnamefont{Warrington}},
  \bibnamefont{and} \bibinfo{author}{\bibfnamefont{D.~C.}
  \bibnamefont{Wilson}}, \bibinfo{journal}{Phys. Rev. A}
  \textbf{\bibinfo{volume}{47}}, \bibinfo{pages}{441} (\bibinfo{year}{1993}),
  \urlprefix\url{https://link.aps.org/doi/10.1103/PhysRevA.47.441}.

\bibitem[{\citenamefont{Raab et~al.}(2000)\citenamefont{Raab, Eschner, Bolle,
  Oberst, Schmidt-Kaler, and Blatt}}]{Raab2000}
\bibinfo{author}{\bibfnamefont{C.}~\bibnamefont{Raab}},
  \bibinfo{author}{\bibfnamefont{J.}~\bibnamefont{Eschner}},
  \bibinfo{author}{\bibfnamefont{J.}~\bibnamefont{Bolle}},
  \bibinfo{author}{\bibfnamefont{H.}~\bibnamefont{Oberst}},
  \bibinfo{author}{\bibfnamefont{F.}~\bibnamefont{Schmidt-Kaler}},
  \bibnamefont{and} \bibinfo{author}{\bibfnamefont{R.}~\bibnamefont{Blatt}},
  \bibinfo{journal}{Phys. Rev. Lett.} \textbf{\bibinfo{volume}{85}},
  \bibinfo{pages}{538} (\bibinfo{year}{2000}),
  \urlprefix\url{https://link.aps.org/doi/10.1103/PhysRevLett.85.538}.

\bibitem[{\citenamefont{Bushev et~al.}(2006)\citenamefont{Bushev, Rotter,
  Wilson, Dubin, Becher, Eschner, Blatt, Steixner, Rabl, and
  Zoller}}]{Bushev2006}
\bibinfo{author}{\bibfnamefont{P.}~\bibnamefont{Bushev}},
  \bibinfo{author}{\bibfnamefont{D.}~\bibnamefont{Rotter}},
  \bibinfo{author}{\bibfnamefont{A.}~\bibnamefont{Wilson}},
  \bibinfo{author}{\bibfnamefont{F.~m.~c.} \bibnamefont{Dubin}},
  \bibinfo{author}{\bibfnamefont{C.}~\bibnamefont{Becher}},
  \bibinfo{author}{\bibfnamefont{J.}~\bibnamefont{Eschner}},
  \bibinfo{author}{\bibfnamefont{R.}~\bibnamefont{Blatt}},
  \bibinfo{author}{\bibfnamefont{V.}~\bibnamefont{Steixner}},
  \bibinfo{author}{\bibfnamefont{P.}~\bibnamefont{Rabl}}, \bibnamefont{and}
  \bibinfo{author}{\bibfnamefont{P.}~\bibnamefont{Zoller}},
  \bibinfo{journal}{Phys. Rev. Lett.} \textbf{\bibinfo{volume}{96}},
  \bibinfo{pages}{043003} (\bibinfo{year}{2006}),
  \urlprefix\url{https://link.aps.org/doi/10.1103/PhysRevLett.96.043003}.

\bibitem[{\citenamefont{Rotter et~al.}(2008)\citenamefont{Rotter, Mukherjee,
  Dubin, and Blatt}}]{rotter2008monitoring}
\bibinfo{author}{\bibfnamefont{D.}~\bibnamefont{Rotter}},
  \bibinfo{author}{\bibfnamefont{M.}~\bibnamefont{Mukherjee}},
  \bibinfo{author}{\bibfnamefont{F.}~\bibnamefont{Dubin}}, \bibnamefont{and}
  \bibinfo{author}{\bibfnamefont{R.}~\bibnamefont{Blatt}},
  \bibinfo{journal}{New Journal of Physics} \textbf{\bibinfo{volume}{10}},
  \bibinfo{pages}{043011} (\bibinfo{year}{2008}),
  \urlprefix\url{https://doi.org/10.1088/1367-2630/10/4/043011}.

\bibitem[{\citenamefont{Bushev et~al.}(2013)\citenamefont{Bushev, H\'etet,
  Slodi\ifmmode~\check{c}\else \v{c}\fi{}ka, Rotter, Wilson, Schmidt-Kaler,
  Eschner, and Blatt}}]{Bushev2013}
\bibinfo{author}{\bibfnamefont{P.}~\bibnamefont{Bushev}},
  \bibinfo{author}{\bibfnamefont{G.}~\bibnamefont{H\'etet}},
  \bibinfo{author}{\bibfnamefont{L.}~\bibnamefont{Slodi\ifmmode~\check{c}\else
  \v{c}\fi{}ka}}, \bibinfo{author}{\bibfnamefont{D.}~\bibnamefont{Rotter}},
  \bibinfo{author}{\bibfnamefont{M.~A.} \bibnamefont{Wilson}},
  \bibinfo{author}{\bibfnamefont{F.}~\bibnamefont{Schmidt-Kaler}},
  \bibinfo{author}{\bibfnamefont{J.}~\bibnamefont{Eschner}}, \bibnamefont{and}
  \bibinfo{author}{\bibfnamefont{R.}~\bibnamefont{Blatt}},
  \bibinfo{journal}{Phys. Rev. Lett.} \textbf{\bibinfo{volume}{110}},
  \bibinfo{pages}{133602} (\bibinfo{year}{2013}),
  \urlprefix\url{https://link.aps.org/doi/10.1103/PhysRevLett.110.133602}.

\bibitem[{\citenamefont{Cerchiari et~al.}(2020)\citenamefont{Cerchiari,
  Araneda, Podhora, Slodička, Colombe, and Blatt}}]{Cerchiari2020}
\bibinfo{author}{\bibfnamefont{G.}~\bibnamefont{Cerchiari}},
  \bibinfo{author}{\bibfnamefont{G.}~\bibnamefont{Araneda}},
  \bibinfo{author}{\bibfnamefont{L.}~\bibnamefont{Podhora}},
  \bibinfo{author}{\bibfnamefont{L.}~\bibnamefont{Slodička}},
  \bibinfo{author}{\bibfnamefont{Y.}~\bibnamefont{Colombe}}, \bibnamefont{and}
  \bibinfo{author}{\bibfnamefont{R.}~\bibnamefont{Blatt}},
  \bibinfo{journal}{arXiv 2009.14098}  (\bibinfo{year}{2020}),
  \urlprefix\url{https://arxiv.org/abs/2009.14098}.

\bibitem[{\citenamefont{Cerchiari et~al.}(2021)\citenamefont{Cerchiari, Dania,
  Bykov, Blatt, and Northup}}]{Cerchiari2021position}
\bibinfo{author}{\bibfnamefont{G.}~\bibnamefont{Cerchiari}},
  \bibinfo{author}{\bibfnamefont{L.}~\bibnamefont{Dania}},
  \bibinfo{author}{\bibfnamefont{D.~S.} \bibnamefont{Bykov}},
  \bibinfo{author}{\bibfnamefont{R.}~\bibnamefont{Blatt}}, \bibnamefont{and}
  \bibinfo{author}{\bibfnamefont{T.}~\bibnamefont{Northup}},
  \bibinfo{journal}{arXiv 2103.08322}  (\bibinfo{year}{2021}),
  \urlprefix\url{https://arxiv.org/abs/2103.08322}.

\bibitem[{\citenamefont{Home et~al.}(2011)\citenamefont{Home, Hanneke, Jost,
  Leibfried, and Wineland}}]{Home_2011}
\bibinfo{author}{\bibfnamefont{J.~P.} \bibnamefont{Home}},
  \bibinfo{author}{\bibfnamefont{D.}~\bibnamefont{Hanneke}},
  \bibinfo{author}{\bibfnamefont{J.~D.} \bibnamefont{Jost}},
  \bibinfo{author}{\bibfnamefont{D.}~\bibnamefont{Leibfried}},
  \bibnamefont{and} \bibinfo{author}{\bibfnamefont{D.~J.}
  \bibnamefont{Wineland}}, \bibinfo{journal}{New Journal of Physics}
  \textbf{\bibinfo{volume}{13}}, \bibinfo{pages}{073026}
  (\bibinfo{year}{2011}),
  \urlprefix\url{https://doi.org/10.1088%2F1367-2630%2F13%2F7%2F073026}.

\bibitem[{\citenamefont{Araneda et~al.}(2018)\citenamefont{Araneda,
  Higginbottom, Slodi{\v{c}}ka, Colombe, and Blatt}}]{Araneda_2018}
\bibinfo{author}{\bibfnamefont{G.}~\bibnamefont{Araneda}},
  \bibinfo{author}{\bibfnamefont{D.~B.} \bibnamefont{Higginbottom}},
  \bibinfo{author}{\bibfnamefont{L.}~\bibnamefont{Slodi{\v{c}}ka}},
  \bibinfo{author}{\bibfnamefont{Y.}~\bibnamefont{Colombe}}, \bibnamefont{and}
  \bibinfo{author}{\bibfnamefont{R.}~\bibnamefont{Blatt}},
  \bibinfo{journal}{Phys. Rev. Lett.} \textbf{\bibinfo{volume}{120}},
  \bibinfo{pages}{193603} (\bibinfo{year}{2018}),
  \urlprefix\url{https://link.aps.org/doi/10.1103/PhysRevLett.120.193603}.

\bibitem[{\citenamefont{Dorner and Zoller}(2002)}]{Dorner2002}
\bibinfo{author}{\bibfnamefont{U.}~\bibnamefont{Dorner}} \bibnamefont{and}
  \bibinfo{author}{\bibfnamefont{P.}~\bibnamefont{Zoller}},
  \bibinfo{journal}{Phys. Rev. A} \textbf{\bibinfo{volume}{66}},
  \bibinfo{pages}{023816} (\bibinfo{year}{2002}),
  \urlprefix\url{https://link.aps.org/doi/10.1103/PhysRevA.66.023816}.

\bibitem[{\citenamefont{H\'etet et~al.}(2010)\citenamefont{H\'etet,
  Slodi\ifmmode~\check{c}\else \v{c}\fi{}ka, Gl\"atzle, Hennrich, and
  Blatt}}]{Hetet2010}
\bibinfo{author}{\bibfnamefont{G.}~\bibnamefont{H\'etet}},
  \bibinfo{author}{\bibfnamefont{L.}~\bibnamefont{Slodi\ifmmode~\check{c}\else
  \v{c}\fi{}ka}}, \bibinfo{author}{\bibfnamefont{A.}~\bibnamefont{Gl\"atzle}},
  \bibinfo{author}{\bibfnamefont{M.}~\bibnamefont{Hennrich}}, \bibnamefont{and}
  \bibinfo{author}{\bibfnamefont{R.}~\bibnamefont{Blatt}},
  \bibinfo{journal}{Phys. Rev. A} \textbf{\bibinfo{volume}{82}},
  \bibinfo{pages}{063812} (\bibinfo{year}{2010}),
  \urlprefix\url{https://link.aps.org/doi/10.1103/PhysRevA.82.063812}.

\bibitem[{\citenamefont{Bushev et~al.}(2004)\citenamefont{Bushev, Wilson,
  Eschner, Raab, Schmidt-Kaler, Becher, and Blatt}}]{Bushev_2004}
\bibinfo{author}{\bibfnamefont{P.}~\bibnamefont{Bushev}},
  \bibinfo{author}{\bibfnamefont{A.}~\bibnamefont{Wilson}},
  \bibinfo{author}{\bibfnamefont{J.}~\bibnamefont{Eschner}},
  \bibinfo{author}{\bibfnamefont{C.}~\bibnamefont{Raab}},
  \bibinfo{author}{\bibfnamefont{F.}~\bibnamefont{Schmidt-Kaler}},
  \bibinfo{author}{\bibfnamefont{C.}~\bibnamefont{Becher}}, \bibnamefont{and}
  \bibinfo{author}{\bibfnamefont{R.}~\bibnamefont{Blatt}},
  \bibinfo{journal}{Phys. Rev. Lett.} \textbf{\bibinfo{volume}{92}},
  \bibinfo{pages}{223602} (\bibinfo{year}{2004}),
  \urlprefix\url{https://link.aps.org/doi/10.1103/PhysRevLett.92.223602}.

\bibitem[{\citenamefont{Ob{\v{s}}il et~al.}(2019)\citenamefont{Ob{\v{s}}il,
  Le{\v{s}}und{\'a}k, Pham, Araneda, {\v{C}}{\'\i}{\v{z}}ek, {\v{C}}{\'\i}p,
  Filip, and Slodi{\v{c}}ka}}]{obvsil2019multipath}
\bibinfo{author}{\bibfnamefont{P.}~\bibnamefont{Ob{\v{s}}il}},
  \bibinfo{author}{\bibfnamefont{A.}~\bibnamefont{Le{\v{s}}und{\'a}k}},
  \bibinfo{author}{\bibfnamefont{T.}~\bibnamefont{Pham}},
  \bibinfo{author}{\bibfnamefont{G.}~\bibnamefont{Araneda}},
  \bibinfo{author}{\bibfnamefont{M.}~\bibnamefont{{\v{C}}{\'\i}{\v{z}}ek}},
  \bibinfo{author}{\bibfnamefont{O.}~\bibnamefont{{\v{C}}{\'\i}p}},
  \bibinfo{author}{\bibfnamefont{R.}~\bibnamefont{Filip}}, \bibnamefont{and}
  \bibinfo{author}{\bibfnamefont{L.}~\bibnamefont{Slodi{\v{c}}ka}},
  \bibinfo{journal}{New Journal of Physics} \textbf{\bibinfo{volume}{21}},
  \bibinfo{pages}{093039} (\bibinfo{year}{2019}),
  \urlprefix\url{https://doi.org/10.1088/1367-2630/ab4081}.

\bibitem[{\citenamefont{Araneda et~al.}(2020)\citenamefont{Araneda, Cerchiari,
  Higginbottom, Holz, Lakhmanskiy, Ob{\v{s}}il, Colombe, and
  Blatt}}]{araneda2020panopticon}
\bibinfo{author}{\bibfnamefont{G.}~\bibnamefont{Araneda}},
  \bibinfo{author}{\bibfnamefont{G.}~\bibnamefont{Cerchiari}},
  \bibinfo{author}{\bibfnamefont{D.~B.} \bibnamefont{Higginbottom}},
  \bibinfo{author}{\bibfnamefont{P.}~\bibnamefont{Holz}},
  \bibinfo{author}{\bibfnamefont{K.}~\bibnamefont{Lakhmanskiy}},
  \bibinfo{author}{\bibfnamefont{P.}~\bibnamefont{Ob{\v{s}}il}},
  \bibinfo{author}{\bibfnamefont{Y.}~\bibnamefont{Colombe}}, \bibnamefont{and}
  \bibinfo{author}{\bibfnamefont{R.}~\bibnamefont{Blatt}},
  \bibinfo{journal}{Review of Scientific Instruments}
  \textbf{\bibinfo{volume}{91}}, \bibinfo{pages}{113201}
  (\bibinfo{year}{2020}), \urlprefix\url{https://doi.org/10.1063/5.0020661}.

\bibitem[{\citenamefont{Lechner et~al.}(2016)\citenamefont{Lechner, Maier,
  Hempel, Jurcevic, Lanyon, Monz, Brownnutt, Blatt, and
  Roos}}]{lechner2016electromagnetically}
\bibinfo{author}{\bibfnamefont{R.}~\bibnamefont{Lechner}},
  \bibinfo{author}{\bibfnamefont{C.}~\bibnamefont{Maier}},
  \bibinfo{author}{\bibfnamefont{C.}~\bibnamefont{Hempel}},
  \bibinfo{author}{\bibfnamefont{P.}~\bibnamefont{Jurcevic}},
  \bibinfo{author}{\bibfnamefont{B.~P.} \bibnamefont{Lanyon}},
  \bibinfo{author}{\bibfnamefont{T.}~\bibnamefont{Monz}},
  \bibinfo{author}{\bibfnamefont{M.}~\bibnamefont{Brownnutt}},
  \bibinfo{author}{\bibfnamefont{R.}~\bibnamefont{Blatt}}, \bibnamefont{and}
  \bibinfo{author}{\bibfnamefont{C.~F.} \bibnamefont{Roos}},
  \bibinfo{journal}{Phys. Rev. A} \textbf{\bibinfo{volume}{93}},
  \bibinfo{pages}{053401} (\bibinfo{year}{2016}),
  \urlprefix\url{https://link.aps.org/doi/10.1103/PhysRevA.93.053401}.

\end{thebibliography}

\section{Appendix}

\begin{figure}
\centering
\includegraphics[width=1\columnwidth]{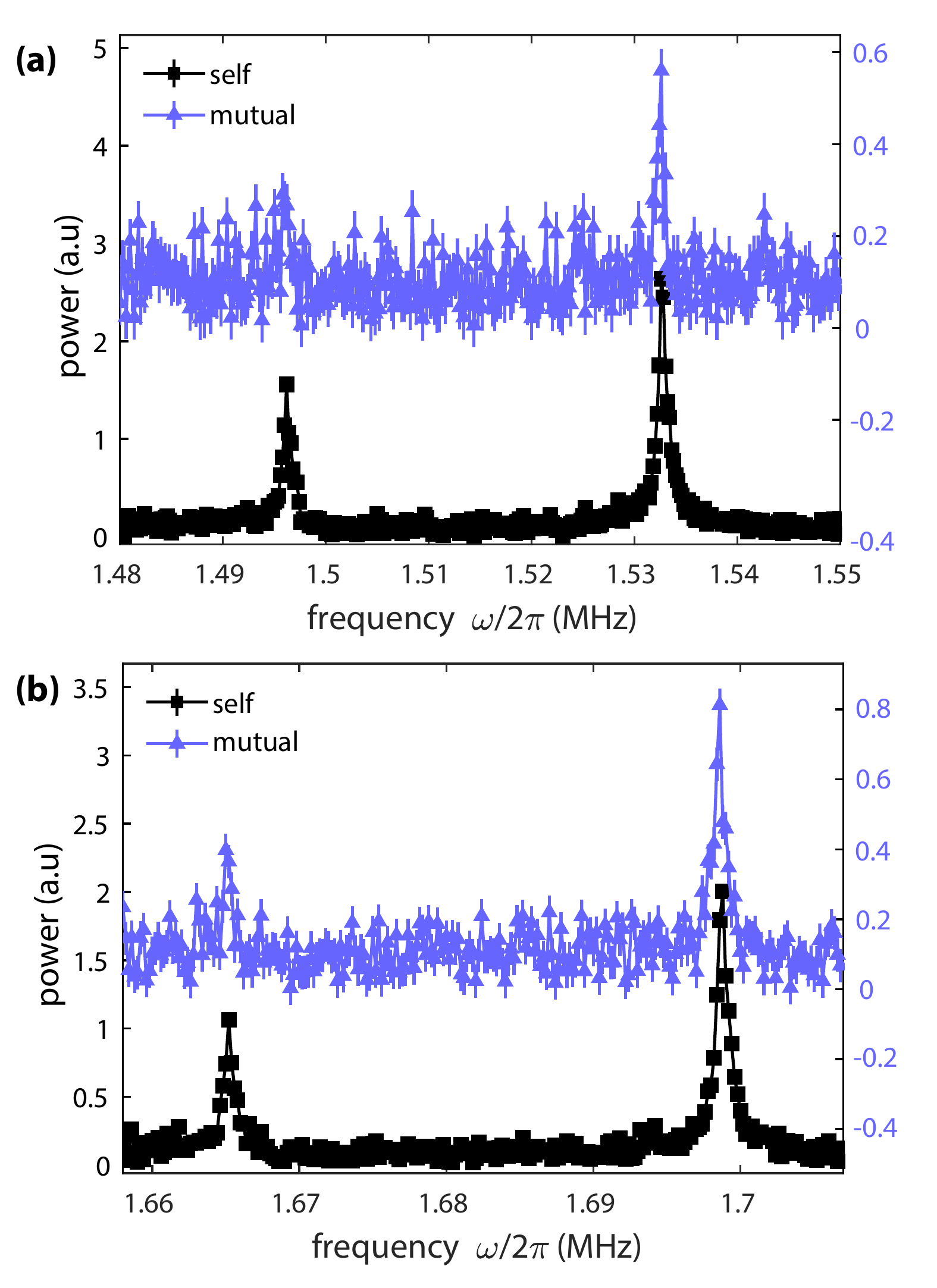}
\caption{Measurements of the ions' radial modes of oscillation under continuous Doppler cooling. (a) Radial rocking modes. (b) Radial common modes. The legend refers to the ``\textit{self}-interference'' and ``\textit{mutual} interference'' configurations. The $y$-scales of the plot are color-coded according to the legend. The lines are meant to guide the eye.}
\label{fig:radial_compare}
\end{figure}

In this appendix, we present the measurements of radial modes of a two-ion chain in ``self-interference'' and ``mutual interference'' configurations (Fig.~1 (b-c) - main text). The measurements are acquired in the continuous regime following the procedure described in the main text. Figure~\ref{fig:radial_compare} presents the data of the radial common and rocking modes measured in the two configurations. Comparing the sensitivity of the two measurements on the same modes, we see that the ``self-interference'' configuration delivers a signal 2 to 5 times stronger than the ``mutual interference'' measurement.

\end{document}